\documentclass[12pt]{article}
\usepackage{amssymb,amsmath,epsfig}

\begin{document}
\title{\bf Gravitational Binding Energy in Charged Cylindrical Symmetry}

\author{M. Sharif \thanks{msharif.math@pu.edu.pk} and M. Zaeem Ul Haq Bhatti
\thanks{mzaeem.math@gmail.com}\\
Department of Mathematics, University of the Punjab,\\
Quaid-e-Azam Campus, Lahore-54590, Pakistan.}

\date{}

\maketitle
\begin{abstract}
We consider static cylindrically symmetric charged gravitating
object with perfect fluid and investigate the gravitational binding
energy. It is found that only the localized part of the mass
function provides the gravitational binding energy, whereas the
non-localized part generated by the electric coupling does not
contribute for such energy.
\end{abstract}

\section{Introduction}

In relativistic astrophysics, the energy required to break up the
body into space is the gravitational binding energy which is found
to be equal to the negative of the gravitational potential energy
under certain conditions. In general relativity, different forms of
mass energy are mixed with gravitational binding energy in implicit
ways that clearly show the relation between binding energy and mass
function. As
\begin{equation*}
\textrm{Binding energy}=\textrm{mass change}\times c^2
\end{equation*}
If we remove the binding energy from the system, it would cause the
removal of mass.

The concept of binding energy is useful for understanding a
gravitational system and also has its importance in the analysis of
systems consisting of charged particles. Gravitational binding
energy of a celestial body is the energy required to expand the
material to infinity. In general relativity, gravitational binding
energy might stabilize a pure electromagnetic object. For an object
to be bound gravitationally, it must contain enough perfect fluid to
generate binding energy which is sufficient to overcome the
electrostatic repulsion.

The self-gravitating energy of a compact object emerges due to its
self-interactions. Misner and Sharp \cite{1} argued that energy is
localized in spherical system. Bondi \cite{2} favored this opinion
and provided the mass of cylindrical systems. Herrera and Di Prisco
\cite{3} found an explicit expression for the active gravitational
mass which has a parameter established through the composition of
relaxation time, energy density, temperature and pressure. Zhang
{\it et al}. \cite{5} explored the effect of charge on the critical
mass of static spherical system and found that the presence of net
charge does not effect the existence of critical mass. Herrera {\it
et al}. \cite{4} evaluated the mass function for the cylindrical
system based on the spherically symmetric case and showed that how
the thermoinertial effects reduce the inertial mass. Recently, we
have found the mass function for the charged cylindrical system and
expressed it in terms of structure scalars \cite{4a}.

Sharif and his collaborators \cite{7} investigated the effects of
electromagnetic field on different aspects of static and non-static
symmetric models. Tiwari {\it et al}. \cite{9} proved that the mass
energy density and pressure of the fluid have the electromagnetic
origin. Using inertial mass with Lorentz approach, Gautreau
\cite{10} provided the correspondence of the electron mass with its
gravitational mass in spherical symmetry. Ohanian \cite{11} proved
that the gravitational and inertial masses are equal for any
arbitrary system of matter and gravitational field under certain
conditions.

Recently, Corne and his collaborators \cite{12} have investigated
the electric coupling and gravitational binding of spherically
symmetric charged objects for both isotropic and anisotropic fluids.
This paper investigates charged isotropic perfect fluid in the
cylindrical symmetry. We evaluate total mass of a charged object and
show that the gravitational binding energy is completely destroyed
if the mass density is removed from the fluid.

\section{Cylindrical Symmetry and Electromagnetic Field}

The static cylindrically symmetric spacetime has the following form
\cite{13}
\begin{equation}\label{1}
ds^2=-A^2dt^{2}+B^2dr^{2}+C^2{d\theta^{2}}+ dz^2.
\end{equation}
Here, $A$ and $B$ are dimensionless while $C$ has the dimension of
$r$. In order to preserve the cylindrical symmetry, we assume the
following constraints
\begin{equation*}
-\infty{\leq}t{\leq}\infty,\quad 0\leq{r}\leq\infty,\quad
0\leq{\theta}\leq{2\pi}, \quad -\infty<{z}<{\infty},
\end{equation*}
on the coordinates. Also, we consider the collapsing cylinder filled
with the charged perfect fluid whose energy-momentum tensor is
defined as
\begin{equation}\label{2}
T_{\alpha\beta}=(\mu+P)V_{\alpha}
V_{\beta}+Pg_{\alpha\beta}+\frac{1}{4\pi}\left(F^{\gamma}_{\alpha}
F_{\beta\gamma}-\frac{1}{4}F_{\gamma\delta}F^{\gamma\delta}g_{\alpha\beta}\right),
\end{equation}
where $\mu,~P$ and $V^\alpha$ are the energy density, pressure and
four velocity of the fluid, respectively and $F_{\alpha\beta}$ is
the Maxwell field tensor. In comoving coordinates, we have
$V_{\alpha}=-A\delta^{0}_{\alpha}$ satisfying
$V^{\alpha}V_{\alpha}=-1.$

The $00$-component of the Einstein tensor is
\begin{equation}\label{3}
G_{00}=\left(\frac{A}{B}\right)^2\left(\frac{C''}{C}-\frac{B'C'}{BC}\right).
\end{equation}
The electric field $\mathbf{E}=E(r)\mathbf{e_{r}}$ is a physical
field (radial electric field), where $\mathbf{e_{r}}$ is the unit
normal vector in the radial direction. The ${00}$-component of the
energy-momentum tensor in electromagnetic field is
\begin{equation}\label{4}
T_{00}=\left(\mu+\frac{E^2}{8\pi}\right)A^2,
\end{equation}
where $F^{0r}$ is taken to be the radial electric field $E^r$. Using
Eqs.(\ref{3}) and (\ref{4}), we can write the ${00}$-component of
the Einstein-Maxwell field equation, $G_{\alpha\beta}=8\pi
T_{\alpha\beta}$, as
\begin{equation}\label{5}
\left(\frac{1}{B}\right)^2\left(\frac{C''}{C}-\frac{B'C'}{BC}\right)=8\pi\mu+{E^2}.
\end{equation}
The electric field is determined by the charge distribution through
Gauss law
\begin{equation}\label{6}
\nabla.\mathbf{E}=4\pi\rho,
\end{equation}
where $\rho$ is the charge density.

To calculate the electric field inside the star, we use divergence
of the electromagnetic field tensor $F^{\mu\nu}_{~~;\nu}=4\pi
J^\mu$, which implies that $E^r_{~;r}=F^{0r}_{~~;r}=4\pi J^0$, where
\begin{equation*}
F^{0r}_{~~;r}=\frac{1}{\sqrt{-g}}\left[\sqrt{-g}F^{0r}\right]_{,r}.
\end{equation*}
Also, $J^\mu$ is the four current whose non-zero component is
$J^0=\rho(r)V^0=\rho(r)A^{-1}$. Thus we have
\begin{equation}\label{7}
F^{0r}_{~~;r}=\frac{1}{\sqrt{-g}}[\sqrt{-g}F^{0r}]_{,r}=4\pi\rho
A^{-1},
\end{equation}
where $\sqrt{-g}=ABC$. Also,
\begin{equation}\label{8}
\frac{1}{\sqrt{-g}}[\sqrt{-g}F^{0r}]_{,r}=\frac{1}{ABC}[ABCF^{0r}]_{,r}
=\frac{1}{ABC}[CF^{\hat{0}\hat{r}}]_{,r},
\end{equation}
where hat indicates unit vector. Equations (\ref{7}) and (\ref{8})
lead to
\begin{equation}\label{9}
\frac{d}{dr}[CF^{\hat{0}\hat{r}}]=4\pi\rho BC.
\end{equation}
Integration of this equation yields
\begin{equation*}
CF^{\hat{0}\hat{r}}=4\pi\int^r_{0}{\rho}BCdr,
\end{equation*}
which can be written as
\begin{equation}\label{10}
CF^{\hat{0}\hat{r}}=q(r),
\end{equation}
where
\begin{equation}\label{11}
q(r)=4\pi\int^r_{0}{\rho}BCdr,
\end{equation}
is the charge contained in the coordinate system of radius $r$
\cite{15}. We can also write Eq.(\ref{10}) as
\begin{equation}\label{12}
F^{\hat{0}\hat{r}}=\frac{q(r)}{C}=E(r).
\end{equation}

\section{Mass Function}

Thorne \cite{16} defined C-energy as the gravitational mass per unit
specific length of the cylinder given by
\begin{equation*}
E(r)=\tilde{m}(r)=\frac{1}{8}(1-l^{-2}{\nabla}^{\alpha}\tilde{r}{\nabla}_{\alpha}\tilde{r}),
\end{equation*}
where $\varrho,~l$ and $\tilde{r}$ are the circumference radius,
specific length and areal radius, respectively, which satisfy
\begin{equation*}
\varrho^2={\xi_{(1)a}}{\xi^a_{(1)}},\quad
l^2={\xi_{(2)a}}{\xi^a_{(2)}},\quad \tilde{r}=\varrho{l}.
\end{equation*}
Here, $\xi_{(1)}=\frac{\partial}{\partial{\theta}}$ and
$\xi_{(2)}=\frac{\partial}{\partial{z}}$ are the Killing vectors in
cylindrical system. The specific energy of the cylinder in the
interior region is written as
\begin{equation}\label{12a}
\tilde{m}(r)=\frac{1}{8}\left[1-\left(\frac{C'}{B}\right)^2\right].
\end{equation}
Differentiating this equation with respect to $r$, we obtain
\begin{equation}\label{13}
\frac{d\tilde{m}}{dr}=-\frac{CC'}{4}\left[\frac{C''}{B^2C}-\frac{B'C'}{B^3C}\right].
\end{equation}
Using Eqs.(\ref{5}) and (\ref{12}) in (\ref{13}), it follows that
\begin{equation}\label{14}
\frac{d\tilde{m}}{dr}=-\left[2\pi{\mu}CC'+\frac{q^2C'}{4C}\right].
\end{equation}

The densities $\mu(r)$ and $\rho(r)$ vanish at the boundary of the
star $r=R$, while the external region of the cylinder,
$r{\geqslant}R$, provides the constant charge $q(r)$
\begin{equation}\label{15}
q(r)=q(R)=Q=4\pi\int^R_{0}{\rho}BCdr.
\end{equation}
For $\mu(r)=0$ and $q(r)=Q$, Eq.(\ref{14}) becomes
\begin{equation}\label{16}
\frac{d\tilde{m}}{dr}=-\frac{Q^2C'}{4C},
\end{equation}
after integration, it turns out to be
\begin{equation}\label{17}
\tilde{m}(r)=-\frac{Q^2}{4}\ln(CN),
\end{equation}
where $N$ is any arbitrary integration constant with dimension of
$1/L$. Also, Eq.(\ref{12a}) yields
\begin{equation}\label{18}
B^2=\frac{C'^2}{1-8\tilde{m}}=\frac{C'^2}{1+2Q^2\ln(CN)}.
\end{equation}

We define a new function $m(r)$
\begin{equation}\label{19}
\tilde{m}(r)=m(r)-\frac{q^2(r)\ln(CN)}{4},
\end{equation}
which leads Eq.(\ref{14}) to
\begin{equation}\label{20}
\frac{d}{dr}\left(m(r)-\frac{q^2(r)\ln(CN)}{4}\right)
=-\left[2\pi{\mu}CC'+\frac{q^2C'}{4C}\right].
\end{equation}
Outside the star, this equation has total correspondence with our
previous assumptions and provides $m(r)=m(R)=constant$ for
$r{\geqslant}R$. However, inside the star, energy density is not
zero and also charge is not constant. Thus Eq.(\ref{11}) yields
\begin{equation}\label{21}
\frac{dq}{dr}=4\pi{\rho}BC.
\end{equation}
Using Eq.(\ref{20}) in (\ref{21}), it follows that
\begin{equation*}
\frac{dm}{dr}=-(2\pi{\mu}CC'-2\pi\rho{B}Cq\ln(CN)).
\end{equation*}
Integrating this equation, we obtain
\begin{equation*}
m(r)=-2\pi\int^r_{0}{\mu}CC'dr+2\pi\int^r_{0}{q(r)}\rho(r)BC\ln(CN)dr,
\end{equation*}
which is called the mass function. Since the total energy is not
localizable in general relativity, so it cannot be interpreted as
the mass energy inside $r$. The expression
\begin{equation*}
M=m(R)=-\int^R_{0}2\pi{\mu}CC'dr+\int^R_{0}2\pi{q(r)}\rho(r)BC\ln(CN)dr,
\end{equation*}
provides the total gravitational mass of the star. The second term
of this expression shows the electromagnetic contribution in the
system, called the electromagnetic mass of the star. Moreover, this
corresponds to an integral over the proper volume, whereas the first
integral does not have such correspondence.

To make the first term as the integral over the proper volume, we
rewrite $M$ in the form
\begin{equation}\label{22}
M=m(R)=-\int^R_{0}B^{-1}2\pi{\mu}BCC'dr+\int^R_{0}2\pi{q(r)}\rho(r)BC\ln(CN)dr,
\end{equation}
where the integration in both terms is taken over the proper volume.
Based on the analysis of the Keplerian motion of neutral test
particles around the star, the above equation identifies the total
gravitational mass of the star (similar to the electromagnetic mass
of an electron in the Lorentz theory). The first term contains the
factor
\begin{equation*}
B^{-1}=\left(\frac{1-8\tilde{m}}{C'^2}\right)^{\frac{1}{2}}
=\left(\frac{1+2q^2\ln(CN)}{C'^2}\right)^{\frac{1}{2}},
\end{equation*}
which represents the contribution of the gravitational binding
energy to the mass of the charged object. Equation (\ref{22}) can be
rewritten as
\begin{eqnarray}\label{23}
M&=&-\int^R_{0}2\pi{\mu}BCC'dr-\int^R_{0}2\pi{q(r)}\rho(r)BC\ln(CN)dr\nonumber\\
&+&\int^R_{0}(1-B^{-1})2\pi{\mu}BCC'dr.
\end{eqnarray}
This shows that the gravity binds only the localized part of the
mass, while it has no contribution for the non-localized part caused
by the electric coupling. Here the integral
\begin{equation*}
\int^R_{0}(1-B^{-1})2\pi{\mu}BCC'dr
\end{equation*}
is the gravitational binding energy of the object. We see that the
gravitational binding energy is affected by the electric charge
through its contribution in $\tilde{m}$. Moreover, we note that the
cylindrical mass has the contribution of electric field from outside
the cylinder as energy transfer may occur from infinity in the axial
direction.

\section{Conclusions}

We have investigated the mass function by using the Einstein-Maxwell
field equations in cylindrical system with perfect fluid. This
electromagnetic mass contains all the properties of the inertial
mass and have localized and non-localized interactions. Also, the
gravity holds only the localized part but does not hold the
non-localized part generated by the electric coupling. When $q(r)=0$
or $\rho=0$, the integral might become negative (23). In this case,
gravity cannot hold the object together and hence the gravitational
binding energy is completely destroyed.


\vspace{0.5cm}

\begin{thebibliography}{40}

\bibitem{1} Misner, C.W. and Sharp, D.H.: Phys. Rev. \textbf{B136}(1964)571.

\bibitem{2} Bondi, H.: Proc. Roy. Soc. London \textbf{A427}(1990)249.

\bibitem{3} Herrera, L. and Di Prisco, A.: Gen. Relativ. Gravit. \textbf{31}(1999)301.

\bibitem{5} Zhang, J.L., Chau, W.Y. and Deng, T.Y.: Astrophys. Space Sci. \textbf{88}(1982)81.

\bibitem{4} Herrera, L., Di
Prisco, A., Ospino, J. and Ib\'{a}\~{n}ez.: Gen. Relativ. Gravit.
\textbf{44}(2012)2645.

\bibitem{4a} Sharif, M. and Bhatti, M.Z.: Gen. Relativ. Gravit.
\textbf{44}(2012)2811.

\bibitem{7} Sharif, M. and Ismat, H.F.: Can. J. Phys.
\textbf{89}(2011)1203; Sharif, M. and Abass, G.: Astrophys. Space
Sci. \textbf{335}(2011)515; Sharif, M. and Siddiqa, A.: Gen.
Relativ. Gravit. \textbf{43}(2011)73; Sharif, M. and Azam, M.: Gen.
Relativ. Gravit. \textbf{44}(2012)1181.

\bibitem{9} Tiwari, R.N., Rao, J.R. and Kanakamedala, R.R.: Phys. Rev. \textbf{D30}(1984)489.

\bibitem{10} Gautreau, R.: Phys. Rev. \textbf{D31}(1985)1860.

\bibitem{11} Ohanian, H.C.: arXiv:1010.5557.

\bibitem{12} Corne, M., Kheyfets, A. and Miller, W.A.: Class. Quantum Grav. \textbf{24}(2007)5999;
Corne, M., Kheyfets, A., Piasio, J. and Voegele, C.: Int. J. Theor.
Phys. \textbf{50}(2011)2737.

\bibitem{13} Sharif, M. and Azam, M.: JCAP \textbf{02}(2012)043.

\bibitem{15} Sharif, M. and Fatima, S.: Gen. Relativ. Gravit. \textbf{43}(2011)127.

\bibitem{16} Thorne, K.S.: Phys. Rev. \textbf{B138}(1965)251.

\end{thebibliography}
\end{document}